\documentclass[10pt,prl,aps,twocolumn,showpacs,amsmath,amssymb,superscriptaddress]{revtex4-1}

\usepackage{amsmath}
\usepackage{amsfonts}
\usepackage{amssymb}
\usepackage{graphicx}
\usepackage{epstopdf}
\usepackage{float}
\usepackage{physics}
\usepackage{psfrag}
\usepackage{bbold}
\usepackage{dcolumn}
\usepackage{xcolor}

\newcommand{\mi}{\mathrm{i}}

\begin{document}
\title{Interaction signatures and non-Gaussian photon states from a strongly driven atomic ensemble coupled to a nanophotonic waveguide}
\author{B. Olmos}
\affiliation{School of Physics and Astronomy and Centre for the Mathematics and Theoretical Physics of Quantum Non-Equilibrium Systems, The University of Nottingham, Nottingham, NG7 2RD, United Kingdom}
\affiliation{Institut f\"ur Theoretische Physik, Universit\"at T\"ubingen, Auf der Morgenstelle 14, 72076 T\"ubingen, Germany}
\author{G. Buonaiuto}
\affiliation{Institut f\"ur Theoretische Physik, Universit\"at T\"ubingen, Auf der Morgenstelle 14, 72076 T\"ubingen, Germany}
\author{P. Schneeweiss}
\affiliation{Department of Physics, Humboldt-Universit\"at zu Berlin, 10099 Berlin, Germany}
\author{I. Lesanovsky}
\affiliation{School of Physics and Astronomy and Centre for the Mathematics and Theoretical Physics of Quantum Non-Equilibrium Systems, The University of Nottingham, Nottingham, NG7 2RD, United Kingdom}
\affiliation{Institut f\"ur Theoretische Physik, Universit\"at T\"ubingen, Auf der Morgenstelle 14, 72076 T\"ubingen, Germany}

\begin{abstract}
We study theoretically a laser-driven one-dimensional chain of atoms interfaced with the guided optical modes of a nanophotonic waveguide. The period of the chain and the orientation of the laser field can be chosen such that emission occurs predominantly into a single guided mode. We find that the fluorescence excitation line shape changes as the number of atoms is increased, eventually undergoing a splitting that provides evidence for the waveguide-mediated all-to-all interactions. Remarkably, in the regime of strong driving the light emitted into the waveguide is non-classical with a significant negativity of the associated Wigner function. We show that both the emission properties and the non-Gaussian character of the light are robust against voids in the atom chain, enabling the experimental study of these effects with present-day technology. Our results offer a route towards novel types of fiber-coupled quantum light sources and an interesting perspective for probing the physics of interacting atomic ensembles through light.
\end{abstract}

\maketitle

\textit{Introduction.} Investigating the interaction of light and matter at the level of single photons and atoms is central to quantum optics. Nanophotonic systems enable strong coupling of emitters to a well-defined optical mode, thereby providing a powerful platform for studying light-matter interaction experimentally \cite{lodahl2015,nieddu2016,solanobook,nayak2018}. In particular, the strong transversal confinement of light in optical nanostructures naturally enables the implementation of chiral light-matter coupling, where the light propagation direction symmetry is broken \cite{lekien2005,mitsch2014,sollner2015,coles2016,mahmoodian2016,lodahl2017,scarpelli2019}. This recently led to the demonstration of, e.g., optical isolators~\cite{Sayrin15b} and circulators~\cite{scheucher2016}. Moreover, this directional coupling is ideally suited for the implementation of so-called cascaded quantum systems~\cite{Carmichael93, Gardiner93} which can show surprising effects such as the formation of entangled spin dimers by driving the system into a steady-state~\cite{stannigel2012,ramos2014,buonaiuto2019}.

Nanophotonic waveguides provide not only strong but also homogeneous coupling, even for large ensembles of emitters, e.g. for optically trapped laser-cooled atoms. The propagation losses of light in nanofiber-based waveguides are, in addition, small \cite{kovalenko2008,hoffman2014}. The light-mediated atom-atom interactions in this system are of extremely long-range character. This enables the theoretical and experimental investigation of many-body-physics with all-to-all interactions \cite{shahmoon2016,kornovan2016,ruostekoski2016,ruostekoski2017,cheng2017,asenjo2017}. Here, the physics depends strongly on the absolute number of constituents which is also the case, for example, for gravitating systems \cite{padmanabhan1990}. Moreover, since these interactions are mediated by the light in the waveguide, the atom-waveguide coupling can be drastically enhanced when a constructive interference condition is met, as demonstrated, e.g., by the observation of large Bragg reflections using ``mirrors'' that consist of only a few hundred atoms~\cite{Corzo16,Soerensen16}.

\begin{figure}[t!]
\includegraphics[width=\columnwidth]{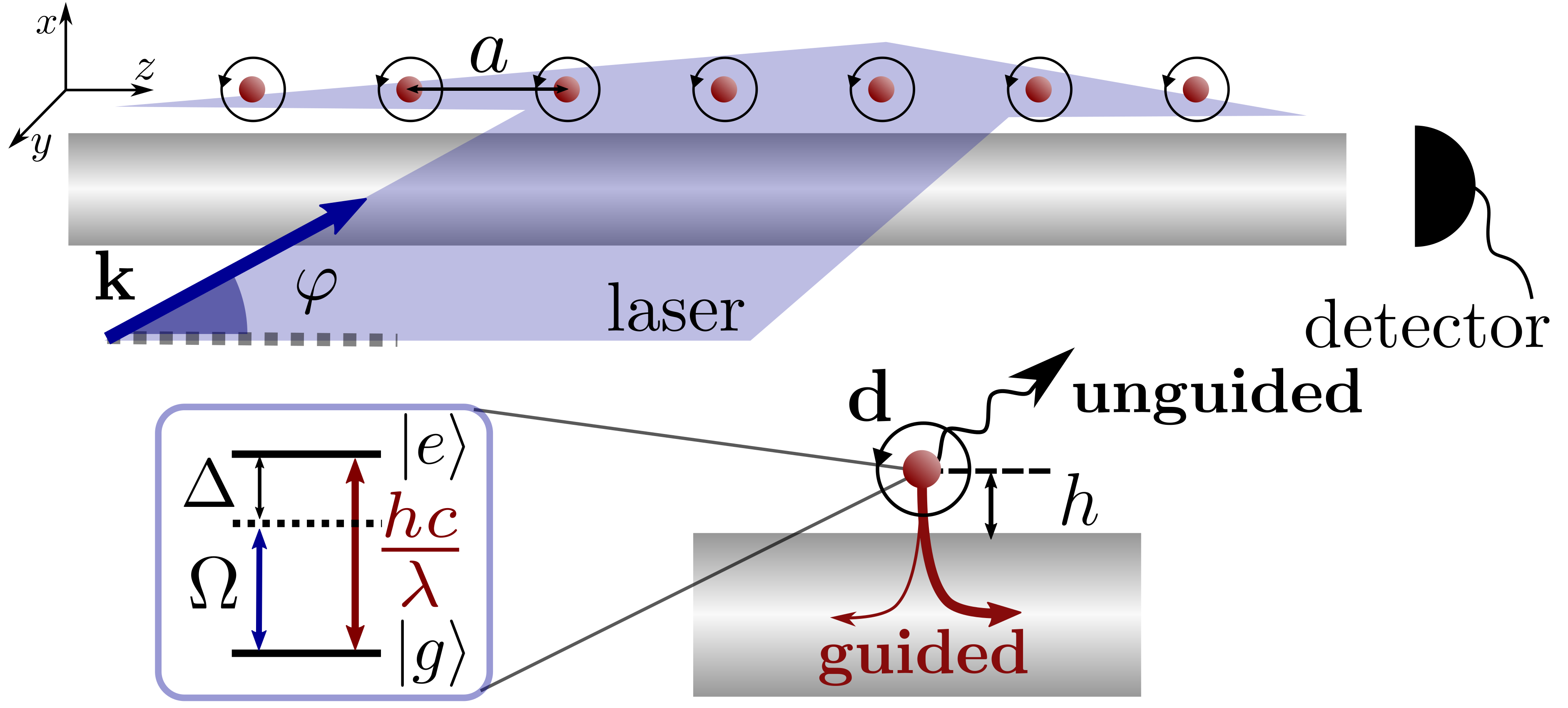}
\caption{\textbf{System.} Periodic array of two-level atoms with nearest neighbor separation $a$, placed at a distance $h$ from a nanofiber. The atomic $\left|g\right>\to\left|e\right>$ transition with dipole moment $\mathbf{d}=d/\sqrt{2}(1,0,-\mi)$ is driven by a laser field with Rabi frequency $\Omega$ and detuning $\Delta$.  Photons emitted in the right-propagating mode are detected.}\label{fig1}
\end{figure}

In this work, we shed light on the signatures of all-to-all interactions in an atomic system that is driven by a laser field and coupled to a nanophotonic waveguide. We operate in the regime of large coupling, where the majority of scattered photons is unidirectionally emitted into the waveguide~\cite{jones2020}. All-to-all interactions manifest in the fluorescence excitation line shape, whose form is strongly particle number dependent. We find a characteristic splitting into a double-peak as the atom number increases. We, moreover, investigate the so far little explored regime of strong driving, where the laser Rabi frequency becomes comparable with the single atom decay rate. Here, the light scattered into the guided mode can become non-classical, which is analyzed through the negativity of the Wigner function of the output light field. This negativity exhibits, as well, a strong particle number dependence, but also shows a high degree of robustness against voids in the atomic chain due to the peculiar nature of the all-to-all interactions. Our study shows that strong driving may enable the controlled realization of non-Gaussian quantum states of light with a potential use as quantum resources \cite{Eis,Cuff,howard2014}.

\textit{System and model.} We consider a periodic array of $N$ atoms with nearest neighbor separation $a$ held at a distance $h$ from the surface of a cylindrical nanofiber (see Fig. \ref{fig1}). We model the internal structure of each atom as a two-level system with states $\left|g\right>$ and $\left|e\right>$. The wavelength of the $\left|g\right>\to\left|e\right>$ transition, $\lambda$, and the radius and dielectric constant of the nanofiber are chosen such that it can only support two pairs of counter-propagating guided modes \cite{snyder1983,malitson1965}. The dipole moment $\mathbf{d}=d/\sqrt{2}(1,0,-\mi)$ and the position of the atoms relative to the nanofiber ensure that the atoms emit only into one of the pairs of modes. The emission is chiral, i.e., each atom has a larger probability to emit, e.g., into the right- than the left-propagating guided mode (accessible with current experimental setups \cite{meng2018}). Moreover, the atoms are coupled to the radiation (unguided) modes of the free electromagnetic field. An external laser field drives the $\left|g\right>\to\left|e\right>$ transition with Rabi frequency $\Omega$ and detuning $\Delta$, and the wave vector of the laser, $\mathbf{k}$, with $k=\left|\mathbf{k}\right|=2\pi/\lambda$, forms an angle $\varphi$ with respect to the atomic chain (as depicted in Fig. \ref{fig1}).

The dynamics of the system is described by the quantum master equation (within the Born-Markov and secular approximations)
\begin{equation}
\dot{\rho}(t)\!=\!-\frac{\mi}{\hbar}\left[H,\rho\right]\!+\!\sum_{ij}\Gamma_{ij}\!\left(\!\sigma_i\rho\sigma_j^\dagger-\frac{1}{2}\left\{\sigma_j^\dagger \sigma_i,\rho\right\}\!\right)\label{mast1}
\end{equation}
with $\sigma_i=\left|g_i\right>\left<e_i\right|$, where $\left|g_i\right>$ and $\left|e_i\right>$ are the ground and excited state of the $i$-th atom in the chain, respectively \cite{lehmberg1970,lekien2017}. The Hamiltonian can be split as $H=H_\mathrm{l}+H_\mathrm{int}$, where the first part accounts for the action of the external laser field
\begin{equation}
H_\mathrm{l}=\hbar\sum_{j=1}^{N}\left(\Omega\left[e^{\mi k a (j-1)\cos{\varphi}}\sigma_j+\text{h.c.}\right]-\Delta\sigma_j^\dagger\sigma_j\right),
\end{equation}
and the second term describes the interatomic interactions due to the exchange of virtual photons among the atoms,
\begin{equation}
H_\mathrm{int}=\hbar\sum_{i\neq j}V_{ij}\sigma_i^\dagger\sigma_j.
\end{equation}
The coherent interaction matrix coefficients $V_{ij}$ can be decomposed into the contributions of the unguided (u) and the right- and left-propagating guided modes (R,L) as $V_{ij}=V_{ij}^\mathrm{u}+V_{ij}^\mathrm{R}+V_{ij}^\mathrm{L}$. While $V_{ij}^\mathrm{u}$ decays with the distance between the atoms in the chain, the guided modes give rise to all-to-all coherent interactions. Similarly, the incoherent emission, which is determined by the matrix coefficients $\Gamma_{ij}$, can be decomposed as $\Gamma_{ij}=\Gamma_{ij}^\mathrm{u}+\Gamma_{ij}^\mathrm{R}+\Gamma_{ij}^\mathrm{L}$. Again here, the guided terms $\Gamma_{ij}^\mathrm{R}$ and $\Gamma_{ij}^\mathrm{L}$ have all-to-all connectivity, which gives rise to collective emission properties largely independent of the nearest neighbor distance $a$ \cite{lekien2017,solano2017,asenjo2017}. On the other hand, $\Gamma_{ij}^\mathrm{u}$ decays with the distance between the atoms, such that the atoms emit independently into the unguided modes as the ratio $a/\lambda$ is increased \cite{lehmberg1970}. The precise expressions of these coefficients can be found in Appendix A.

\begin{figure}[t!]
\includegraphics[width=\columnwidth]{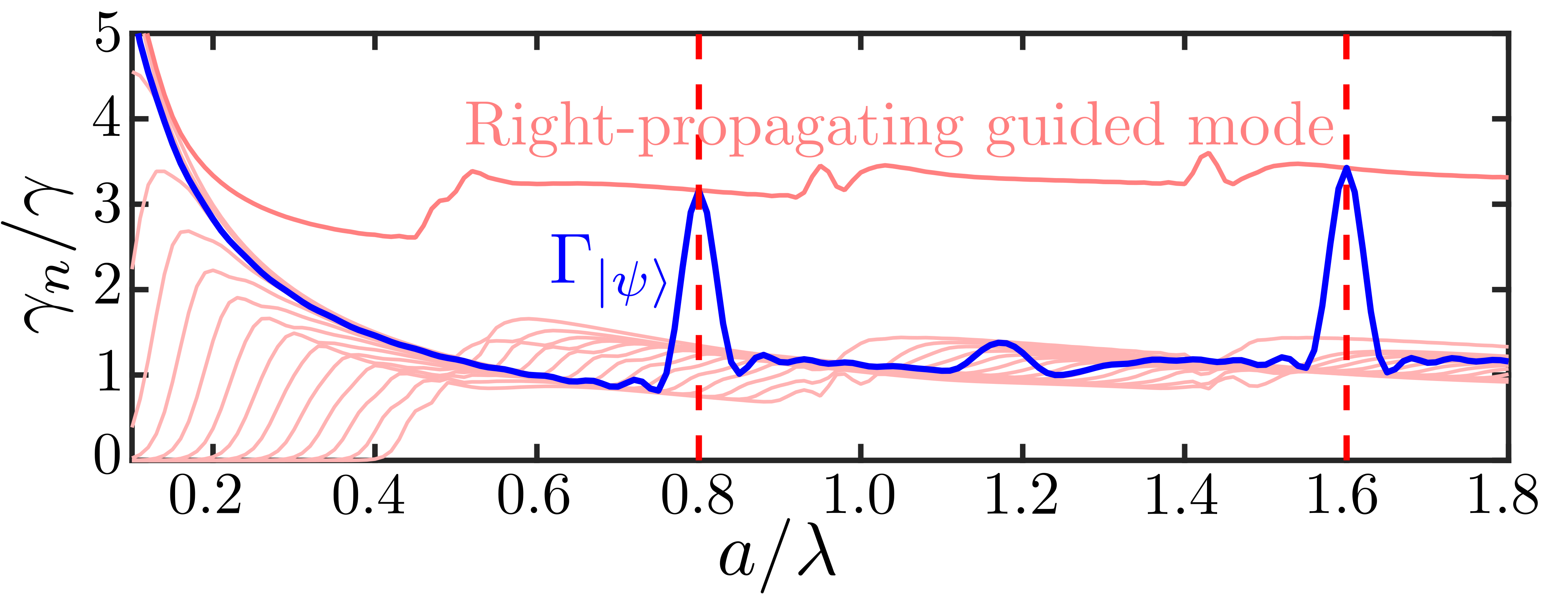}
\caption{\textbf{Effective decay rate of the laser excited state.} Collective decay rates (pink lines), $\gamma_n$ (divided by the single-atom decay rate in the absence of the nanofiber, $\gamma$), for a chain of $N=15$ atoms with $\lambda=1$ $\mu$m. The chain is at a distance $h=100$ nm from a silica nanofiber with refractive index 1.45 and radius $220$ nm, such that the single-atom beta factor (ratio between the emission to the guided modes and the total emission) is $0.15$. The laser momentum forms an angle $\varphi=1.37$ rad with the chain (see Fig. \ref{fig1}). At the values of $a/\lambda$ that satisfy (\ref{eq:fit}) (indicated by the vertical red dashed lines), the effective decay rate $\Gamma_{\left|\psi\right>}$ (blue solid line) matches the one of the right-propagating guided mode.}\label{fig2}
\end{figure}

\textit{Enhanced atom-nanofiber coupling.} In Ref. \cite{jones2020} it was shown that emission takes place predominantly into a single guided mode when the distance between neighboring atoms $a$ and the driving angle $\varphi$ satisfy the relation
\begin{equation}\label{eq:fit}
\cos{\varphi}=\frac{m\lambda}{a}-\frac{\lambda}{\lambda_\mathrm{f}},\qquad m\in \mathbb{N}
\end{equation}
with $\lambda_\mathrm{f}$ being the wavelength of the light guided by the nanofiber. Under this matching condition, not only is the efficiency of the coupling between the atomic array and the nanofiber collectively enhanced \cite{lekien2008,jones2020}, but the photon emission also becomes unidirectional. This phase-matching mechanism can be nicely illustrated in the weak driving limit by assuming that the collective state of the atoms which is excited by the laser takes the form
\begin{equation}\label{eq:psi}
\left|\psi\right>=\sum_{j=1}^N \psi_j \left|e_j\right>,
\end{equation}
with $\psi_j=e^{\mi ka(j-1)\cos{\varphi}}/\sqrt{N}$. The effective decay rate of this state can be calculated as $\Gamma_{\left|\psi\right>}=\sum_{n=1}^N \gamma_n |\sum_{j=1}^N D_{nj}\psi_j|^2$, where $\gamma_n$ and $D_{nj}$ are the collective decay rates and modes, respectively, obtained from the diagonalization of the matrix $\Gamma_{ij}$. As shown in Fig. \ref{fig2}, this decay rate matches the maximally achievable collective rate (which, as shown in Ref. \cite{jones2020}, corresponds to the decay rate into the right-propagating mode) when the ratio $a/\lambda$ satisfies Eq. (\ref{eq:fit}) (red dashed lines) \cite{Note1}.

\begin{figure}[t!]
\includegraphics[width=\columnwidth]{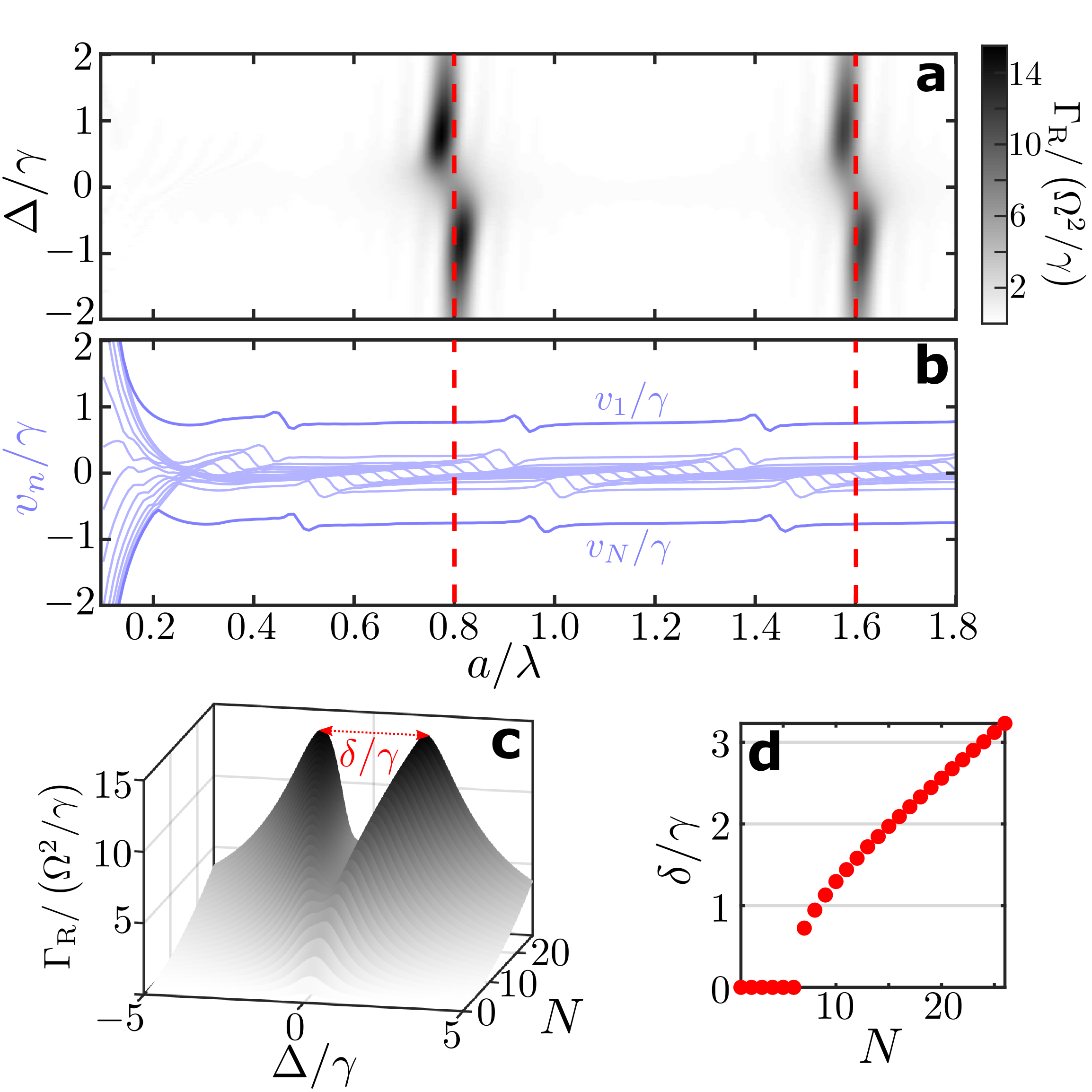}
\caption{\textbf{Line splitting.} \textbf{a:} Emission rate into the right-propagating guided mode, $\Gamma_\mathrm{R}$ (in units of $\Omega^2/\gamma$, total single-atom scattering rate in the weak driving limit), as a function of the laser detuning and the ratio $a/\lambda$ for a laser angle $\varphi=1.37$ rad with the chain, $\Omega=\gamma/100$, and $N=15$. The red dashed lines indicate where condition (\ref{eq:fit}) is satisfied. \textbf{b:} Eigenvalues $v_n$ of the interaction matrix $V_{ij}$. \textbf{c:} Fluorescence excitation line shape, and \textbf{d:} splitting $\delta$ between the two local maxima, as a function of $N$ for $\varphi=1.37$ rad and $a/\lambda=0.8$.}\label{fig3}
\end{figure}

\textit{Interaction signatures at weak driving.} To analyze the properties of the light emitted into the right-propagating mode, we calculate the photon emission rate
\begin{equation}
\Gamma_\mathrm{R}=\sum_{ij}\Gamma_{ij}^\mathrm{R} \left<\sigma_i^\dag\sigma_j\right>_\mathrm{ss},
\end{equation}
where $\left<\dots\right>_\mathrm{ss}$ denotes the expectation value in the stationary state. In the weak driving limit, where at most one excitation is present at all times, this quantity can be calculated for large atom numbers $N$ \cite{jones2020}. In Fig. \ref{fig3}a we show that, indeed, as one approaches the matching condition (\ref{eq:fit}) (red dashed lines), $\Gamma_\mathrm{R}$ quickly becomes maximal, as almost all light is emitted into the fiber. In fact, here $\Gamma_\mathrm{R}\approx N \Omega^2/\gamma$.

Remarkably, the fluorescence excitation line shape changes as the number of atoms increases. Under the phase-matching condition (\ref{eq:fit}) we observe a splitting of the line into two peaks (see Fig. \ref{fig3}c), whose distance $\delta$ increases proportionally to $N$ (for $N$ sufficiently large) as shown in Fig. \ref{fig3}d. This splitting is consequence of the fiber-mediated coherent interactions. This can be better understood by inspecting the eigenvalues $v_n$ of the interaction matrix $V_{ij}$ (Fig. \ref{fig3}b), which represent the energy shift of the corresponding collective eigenstate, $C_{nj}$, with respect to single-atom resonance. At the phase-matching point, the state (\ref{eq:psi}) can be approximately written as an equal weight superposition of the two (guided) extremal eigenmodes $C_{1j}$ and $C_{Nj}$. Their eigenvalues, $v_1$ and $v_N$ (see Fig. \ref{fig3}b), increase linearly with $N$, but with opposite sign. This $N$-dependence is directly reflected in the observed line splitting.

\textit{Strong driving.} For strong driving, more than one atom can be excited at a time. Here, we need to solve the exact master equation, which constrains the number of atoms that we can simulate to a maximum of $N=7$.

\begin{figure}[t]
\label{fig_wig}
\includegraphics[width=\columnwidth]{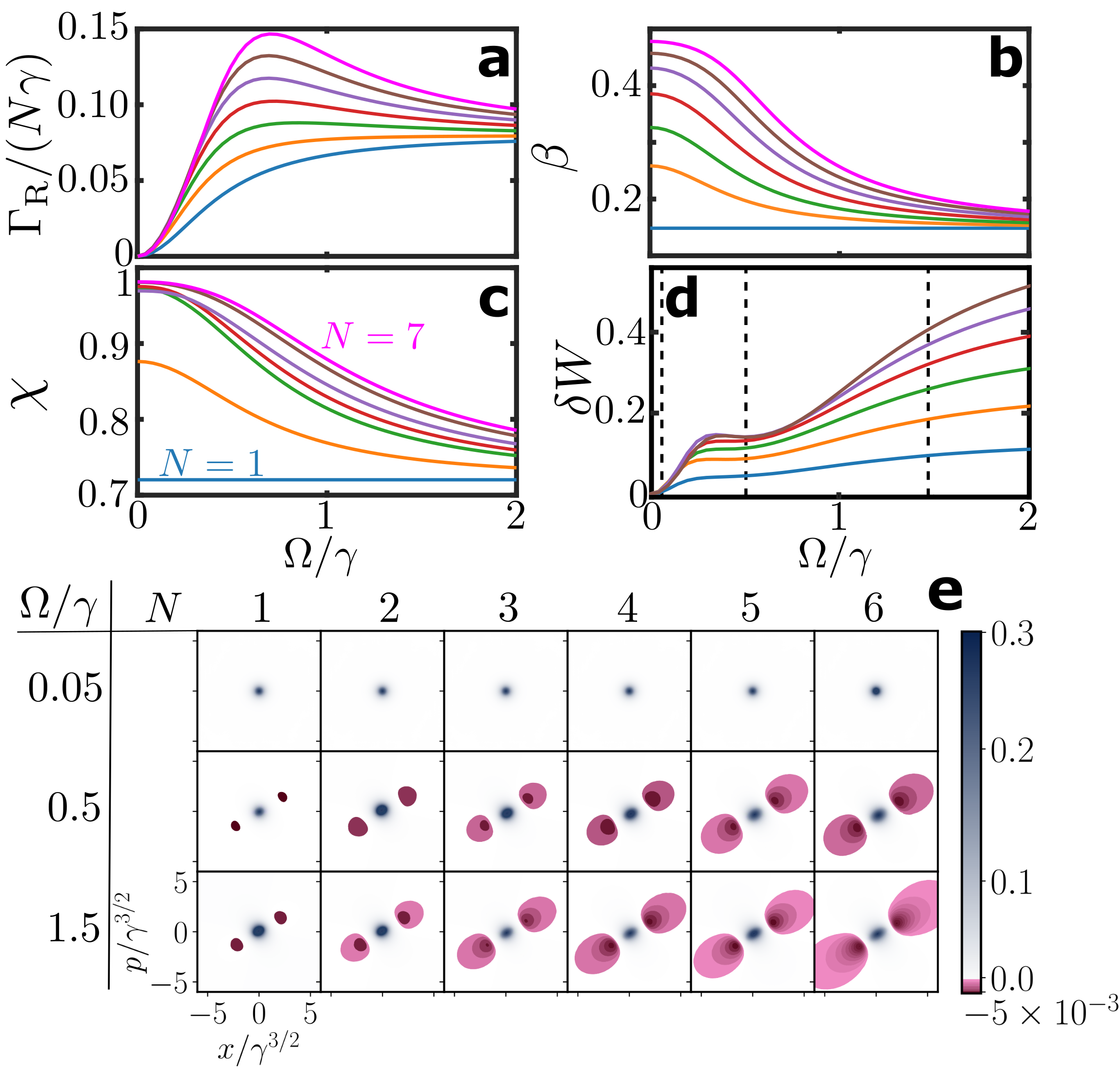}
\caption{\textbf{Strong driving and Wigner distribution.} \textbf{a, b} and \textbf{c:} Emission rate into the right-propagating guided mode, collective beta factor and chirality, respectively (on resonance) for $N=1-7$ atoms. \textbf{d:} Negativity of the WD for $N=1-6$ atoms. \textbf{e:} WD for $\Omega/\gamma=0.05$, $0.5$ and $1.5$. For low values of the driving field, the WD is always positive, regardless of the number of atoms. For $\Omega/\gamma>1$ the negative area of the distribution grows with $N$.}\label{fig4}
\end{figure}

In Fig. \ref{fig4}a, we show the normalized emission rate $\Gamma_\mathrm{R}$ at $\Delta=0$ as a function of the Rabi frequency $\Omega$. If the atoms were non-interacting, all curves would fall into the $N=1$ one. We observe a clear departure from this behavior as $N$ is increased. Deviations are particularly visible when the Rabi frequency is comparable to the single atom decay rate $\gamma$. Here, the normalized emission rate exhibits a local maximum  whose value increases with the number of atoms, indicating that the rate grows faster than linearly with $N$. Similarly, the collective beta factor, $\beta=(\Gamma_\mathrm{R}+\Gamma_\mathrm{L})/(\Gamma_\mathrm{R}+\Gamma_\mathrm{L}+\Gamma_\mathrm{u})$, which quantifies the fraction of the photons that are emitted into the guided modes, as well as the chirality of the photon emission $\chi=(\Gamma_\mathrm{R}-\Gamma_\mathrm{L})/(\Gamma_\mathrm{R}+\Gamma_\mathrm{L})$ (see Figs. \ref{fig4}b and c) increase with the number of atoms and are maximal at weak driving.

\textit{Non-Gaussian state of the guided light.} Signatures of non-classicality can be observed already at intermediate driving strength. To characterise the state of the light emitted from the nanofiber mode, we perform an analysis of the time-integrated photo current observed via homodyne detection $X_\alpha$ \cite{strandberg2019}. For long times, i.e. in the stationary state, the distribution function or marginal, $\Pi_t (X_\alpha)$, of the photo current can be calculated from the largest eigenvalue of a deformed master operator \cite{hickey2012}. In this limit, it obeys a so-called large deviation form, $\Pi_t (X_\alpha )\sim\mathrm{e}^{-t\phi(x_\alpha )}$, with $\phi$ being the rate function and $x_\alpha=X_\alpha/t$ the time-intensive photo current per unit time (activity). At very long times, $\Pi_t (X_\alpha)$ converges to a delta-distribution peaked at the mean photo current per unit time. Quantum features, such as negative Wigner functions, can emerge when the rate function differs from a Gaussian shape. However, these can only be uncovered when the interval of the time integration is not too long, i.e. when $\Pi_t (X_\alpha)$ has not yet converged to a delta-distribution. Integration-time dependent Wigner functions were also observed in \cite{strandberg2019}, where the photo current was calculated from a stochastic master equation. The approach we pursue here allows us to directly access the rate function and thus the distribution function of the photo current without the need to perform Monte Carlo averages. Non-classical features of the output light field are then studied by analysing the marginal distributions $\Pi_{t=1} (X_\alpha )\sim \mathrm{e}^{-\phi(x_\alpha) }$ from which we reconstruct the Wigner distribution (WD) $W(x,p)$ of the $X$- and $P$-quadrature activities $x=X/t$ and $p=P/t$, respectively, using a Radon transform (see Appendix B for more details).

We will quantify the non-Gaussianity of the state of light emitted by the nanofiber through the negativity of the WD \cite{straka2014,happ2018}, which can be calculated as the volume of its negative part, i.e.
\begin{equation}
\label{dw}
\delta W= \int_{\mathbb{R}^2} \left[|W(x,p)|-W(x,p)\right] dx dp.
\end{equation}
A non-zero $\delta W$ is symptomatic of a non-Gaussian state of light, while $\delta W=0$ for coherent and squeezed states \cite{Hud,Zyc}. The negativity, as well as the full WD for selected values of $\Omega/\gamma$ are shown in Fig. \ref{fig4}d and e, respectively. Here, we observe that the negativity tends to zero for small values of $\Omega$, where the response of the system is linear and hence Gaussian states are expected (first row in Fig. \ref{fig4}e). For $\Omega\gg\gamma$, the nonlinear response of the two-level systems leads to negativity in the WD and hence non-gaussianity of the emitted light \cite{John,strandberg2019}, which grows with the number of atoms $N$ (third row in Fig. \ref{fig4}e). For large enough $N$, the negativity $\delta W$ displays non-monotonic behavior and for intermediate values of $\Omega$, as a result of the interplay between the driving and the all-to-all interactions. 

\begin{figure}[t]
\label{fig_wig}
\includegraphics[width=\columnwidth]{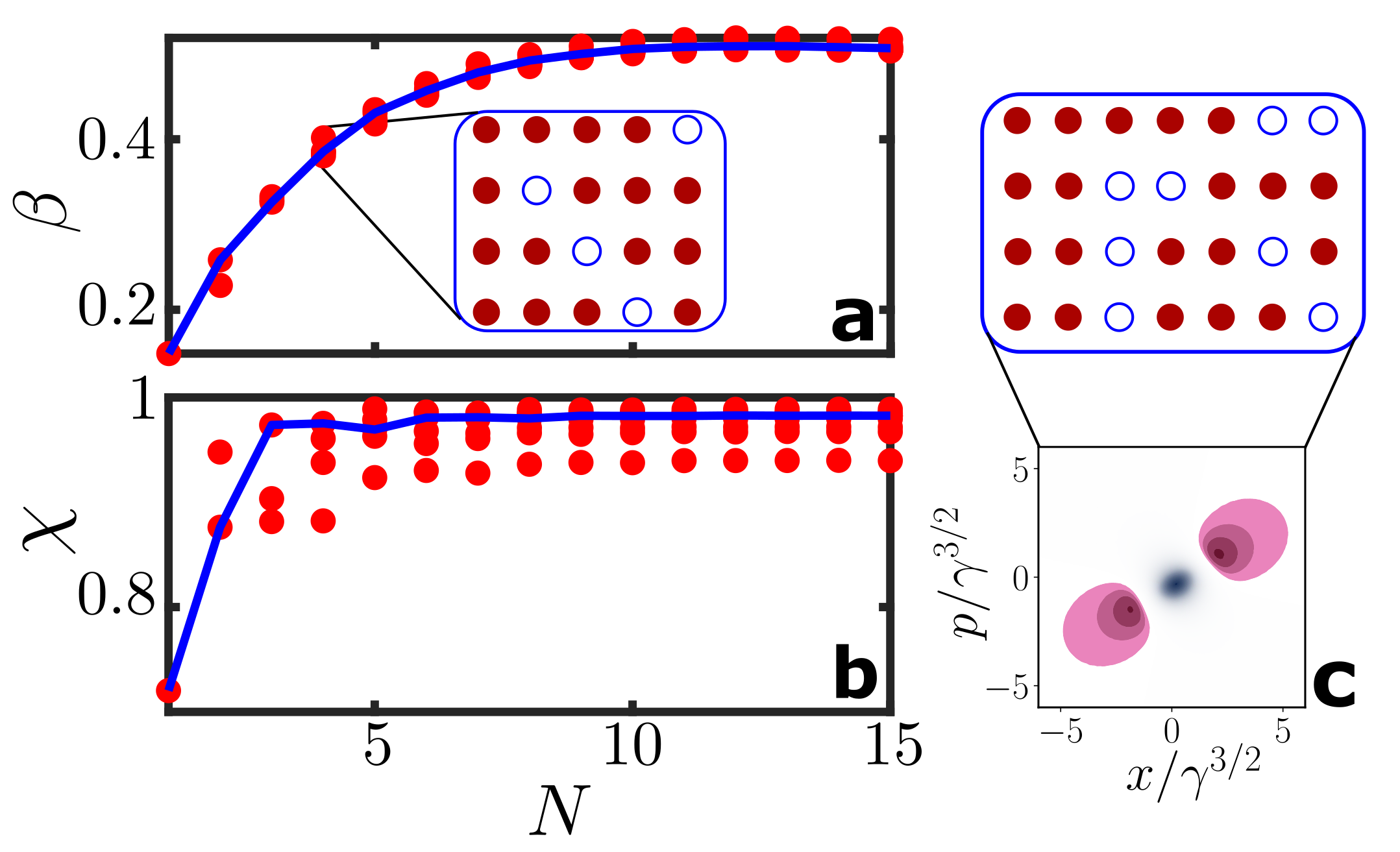}
\caption{\textbf{Robustness to imperfect filling.} \textbf{a:} Collective beta factor and \textbf{b:} chirality of the emission from the weakly driven ($\Omega=\gamma/100$, $\Delta=0$) atomic chain as a function of the system size $N$. The solid lines join the data points for a chain without gaps while the rest of points represent configurations with the same number of atoms but a single gap. \textbf{c:} Four different configurations of $N=5$ atoms in a chain of 7 sites and Wigner distribution at $\Omega=0.5\gamma$, virtually unchanged for all configurations.}\label{fig5}
\end{figure}

\textit{Robustness due to all-to-all interactions}. In a realistic experimental apparatus involving atoms interacting with a nanofiber, often the filling of the traps, created either by the evanescent field of the nanofiber or an external lattice, is imperfect, with gaps appearing in the chain. Here, we demonstrate that the effects described in this work are robust against this experimental imperfection.

In the limit of weak driving, we show in Figs. \ref{fig5}a and b the beta factor and chirality as a function of the number of atoms in the chain $N$, for perfect filling (joined by the solid line) and for different configurations of the same number of atoms leaving a single gap within the chain (see inset in Fig. \ref{fig5}a for all configurations considered for $N=4$). The impact on both $\beta$ and $\chi$ is very small, sometimes even improving on the regularly spaced array. Also the non-Gaussianity of the photon state is robust against imperfect filling. E.g., in Fig. \ref{fig5}c we show the WD, virtually unchanged for several gapped configurations of the same number of atoms $N=5$ in the chain, with differences in the negativity of the order of $10^{-3}$ between different configurations.

We attribute this robustness to the periodicity of the all-to-all interactions between the atoms mediated by the nanofiber: the origin of the collective enhancement of the beta factor and chirality is geometrical and hence any displacement of the atoms will result in a relative phase of $2\pi$, which does not affect the emission properties significantly.

\textit{Conclusions.} We have shown that the light scattered unidirectionally from a laser-driven array of atoms into a nanofiber contains signatures of the all-to-all interactions induced by the atom-fiber coupling. Moreover, in the strong driving regime, we demonstrate a novel realization of a source of quantum non-gaussian states of light in the optical regime. This light can be transported through optical fibers with very small losses for long distances, hence making the findings in this work particularly relevant for optical quantum information and communication. Moreover, it has been shown that a quantum circuit containing operations that generate states represented by a negative Wigner function cannot be efficiently simulated with classical algorithms \cite{Eis,Cuff,howard2014}. Hence, the atom-nanofiber setup considered in this work is a candidate for generating quantum computational resources by making use of state-of-the-art light-matter technologies.

Finally, note that the results we present here can be realized within current experimental capabilities. For example, cesium atoms are routinely trapped close to the surface of a nanofiber, at distances of only about 200 nm \cite{mitsch2014,meng2018}. While these parameters give rise to a weaker single-atom coupling to the nanofiber, the all-to-all character of the interactions induced by the guided modes makes it possible to observe all the effects discussed here by increasing the number of atoms further.

\begin{acknowledgments}
The authors acknowledge fruitful discussions with Christian Liedl, Yijian Meng and Sebastian Pucher. The research leading to these results has received funding from the European Union's H2020 research and innovation programme [Grant Agreement No. 800942 (ErBeStA)] and EPSRC [Grant No. EP/R04340X/1]. IL acknowledges support from the "Wissenschaftler-R\"{u}ckkehrprogramm GSO/CZS“ of the Carl-Zeiss-Stiftung and the German Scholars Organization e.V.. BO was supported by the Royal Society and EPSRC [Grant No. DH130145].
\end{acknowledgments}

\section{Appendix A: Coherent and incoherent interaction matrices: guided and unguided modes}

In this Section we give some details of the steps used for the calculation of the coherent and incoherent interaction terms $V_{ij}$ and $\Gamma_{ij}$, respectively, which appear in the main manuscript.

In order to calculate these coefficients, the key ingredients are the coupling strength functions $G_{\nu i}$ between the atom $i$ and the mode $\nu$ which, for the guided ($\mathrm{g}$) and unguided ($\mathrm{u}$) modes, respectively, are given by \cite{lekien2017}
\begin{align} \label{eq:Gmi}
&G_{\mathrm{g}i} = \sqrt{\frac{\omega \beta_\mathrm{f}^{'}}{4 \pi \hbar \varepsilon_0}} \left[ \mathbf{d} \cdot \mathbf{e}^{(\mathrm{g})}(r_i, \phi_i) \right] \mathrm{e}^{i(f \beta_\mathrm{f} z_i + l \phi_i)}, \nonumber \\
&G_{\mathrm{u}i} = \sqrt{\frac{\omega}{4 \pi \hbar \varepsilon_0}} \left[ \mathbf{d} \cdot \mathbf{e}^{(\mathrm{u})}(r_i, \phi_i) \right] \mathrm{e}^{i(\beta z_i + m \phi_i)},
\end{align}
where $\omega$ is the mode frequency, $l = \pm1$ denotes counterclockwise or clockwise polarization, $f = \pm1$ denotes whether the mode propagates in the $+z$ or $-z$ direction (forwards of backwards along the nanofiber), and $m = 0, \pm 1, \pm 2, ...$ denotes the mode order. $\beta$ is a continuous variable in the range $-k < \beta < k$ with $k = \omega_\mathrm{a} / c$, where $\omega_\mathrm{a}$ is the frequency of the atomic transition. Furthermore, $(r_i,\phi_i,z_i)$ represent the position of the $i$-th atom in cylindrical coordinates (note, that the axis of the nanofiber is aligned along the $z$-axis), $\mathbf{d}$ is the atomic dipole moment, and $\mathbf{e}^{(\nu)}$ denotes the profile function of the electric field. The explicit form of each of these profile functions can be found in the literature, e.g., in \cite{lekien2017}. We consider for our purposes that that the nanofiber only supports a single fundamental pair of guided modes, $\mathrm{HE}_{11}$. Thus, the value of the longitudinal propagation constant of the guided mode $\beta_\mathrm{f}=2\pi/\lambda_\mathrm{f}$ and its derivative $\beta_\mathrm{f}' = \mathrm{d}\beta_\mathrm{f}/\mathrm{d}\omega$, which must be determined numerically as the solution of an eigenvalue equation \cite{snyder1983}, are the only solutions that lie in the range $k < \beta_\mathrm{f} \leq k n_\mathrm{f}$, with $n_\mathrm{f}$ being the dielectric constant of the fiber.

As demonstrated, e.g. in \cite{lekien2017}, the coherent interaction coefficients can be calculated as
\begin{equation}\label{eq:Vab}
V_{ij} = - \mathcal{P} \sum_{\nu} \left[ \frac{G^{}_{\nu i} G^*_{\nu j}}{\omega - \omega_\mathrm{a}} + (-1)^{\delta_{ij}} \frac{\tilde{G}^*_{\nu i} \tilde{G}^{}_{\nu j}}{\omega + \omega_\mathrm{a}} \right],
\end{equation}
and
\begin{equation}\label{eq:Gab}
\Gamma_{ij} = 2 \pi \sum_{\nu} G^{}_{\nu i} G^*_{\nu j} \delta{(\omega - \omega_\mathrm{a})}.
\end{equation}
Here, $\mathcal{P}$ denotes the Cauchy principal value, $\delta_{ij}$ denotes the Kronecker delta function, and the tilde (e.g. $\tilde{G}^{}_{\nu j}$) serves to indicate that the dipole moment $\mathbf{d}$ is to be replaced with its complex conjugate. The sum can be separated into two, over guided and unguided modes $\sum_{\nu} = \sum_\mathrm{g} + \sum_\mathrm{u}$, such that both coherent and incoherent interactions can be separated as $V_{ij}=V_{ij}^\mathrm{g}+V_{ij}^\mathrm{u}$ and $\Gamma_{ij}=\Gamma_{ij}^\mathrm{g}+\Gamma_{ij}^\mathrm{u}$. Moreover, the sum over the guided modes reads $\sum_\mathrm{g} = \int^{\infty}_0 d\omega \sum_{fl}$, such that it can be broken down into the two directions along the fiber, and hence $V_{ij}^\mathrm{g}=V_{ij}^\mathrm{R}+V_{ij}^\mathrm{L}$ and $\Gamma_{ij}^\mathrm{g}=\Gamma_{ij}^\mathrm{R}+\Gamma_{ij}^\mathrm{L}$. For the unguided modes, $\sum_\mathrm{u} = \int^\infty_0 d \omega \int^{k}_{-k} d \beta \sum_{ml}$.

\section{Appendix B: Wigner distribution}

Here we give a detailed explanation on the procedure for evaluating the Wigner distribution (WD) for the time-integrated photo current per time. We analyze the quantum state of the light emitted into the reservoir (here the right-propagating guided mode) by studying the time-integrated quadratures. A similar approach was pursued in \cite{strandberg2019} to analyze resonance fluorescence in the stationary state. The challenge is that the light field at the output is not confined to a cavity with a well-defined stationary quantum state but is rather characterized by continuous-mode field operators $B(t)$. To deal with this situation it is convenient to define time-integrated output modes \cite{strandberg2019}. To this end, we define the time-integrated photo current observed via homodyne detection as $X_\alpha=\int_0^t \mathrm{d}X_\alpha(\tau)$, with
\begin{eqnarray*}
\mathrm{d}X_\alpha (\tau)&=&1/2 \left(\mathrm{e}^{-i\alpha} \mathrm{d}B(\tau)+\mathrm{e}^{i\alpha} \mathrm{d}B(\tau)\right)\\
&=&\mathrm{d}X(\tau)\cos{\alpha}+\mathrm{d}P(\tau) \sin{\alpha}.
\end{eqnarray*}
Here, $\mathrm{d}B(t)$ and $\mathrm{d}B^\dag(t)$ are the increments of the output field operators \cite{hickey2012}, which can be expressed in terms of increments of the $X$- and $P$-quadratures.

We now consider projections of the density matrix $\rho(t)$ onto the subspace for which the time-integrated quadrature for light emitted from the system up to time $t$ has a specific value $X_\alpha$.  We denote the corresponding projected reduced density matrix by ${\rho}^{\left({X}_{\alpha}\right)}_{t}$. This quantity is related to an $s$-biased reduced density matrix ${\rho}^{s}(t)$, via the Laplace transform \cite{Igr,touchette2009},
\begin{equation}\label{eq:7e}
{\rho}^{s}(t) = \int\rho^{\left({X}_{\alpha}\right)}(t){e}^{-s{X}_{\alpha}}\mathrm{d}{X}_{\alpha}.
\end{equation}
We obtain then the moment generating function $Z_{t}(s) = \Tr \rho^{s}(t)$, which at long times takes the \textit{large deviation} (LD) form ${e}^{t\theta_{X_\alpha}(s)}$. The full statistics of each $X_\alpha$ at long times is contained within the scaled cumulant generating functions (SCG) $\theta_{X_{\alpha}}(s)$, which are also identified as the largest real eigenvalue of the deformed master equation \cite{hickey2012}:
\begin{eqnarray}
{\dot{\rho}}^{s}(t)=&&-\frac{\mi}{\hbar}\left[H,\rho^{s}\right]\!+\!\sum_{ij}\Gamma_{ij}\!\left(\!\sigma_i\rho^{s}\sigma_j^\dagger-\frac{1}{2}\left\{\sigma_j^\dagger \sigma_i,\rho^{s}\right\}\!\right) \nonumber \\
&& -\frac{s}{2}\left({e}^{-\mi\alpha}{J_\mathrm{R}}{\rho}^{s}+{e}^{\mi\alpha}{\rho}^{s}{J_\mathrm{R}}^{\dag}\right)+\frac{{s}^{2}}{8}{\rho}^{s}\label{sup2}
\end{eqnarray} 
where $J_\mathrm{R}=\sum_{j}\sqrt{\Gamma^\mathrm{R}_{jj}}e^{-\mi\frac{2 \pi}{\lambda_\mathrm{f}}a(j-1)}\sigma_{j}$, is the jump operator associated with the right-propagating guided mode in the nanofiber.  When $s \to 0$, Eq. \eqref{sup2} collapses to the standard trace-preserving master equation [Eq. (1) in main manuscript]. Away from $s=0$ the $s$-field biases the dynamics towards rare events in the measured light field. 
From the projected reduced density matrix ${\rho}^{\left({X}_{\alpha}\right)}_{t}$ it is possible to evaluate the marginal probability to observe a particular value of $X_{\alpha}$ as $\Pi_t(X_\alpha)=\mathrm{Tr}\left[\rho^{(X_\alpha)}(t)\right]$. At long times (stationary state) this marginal probability takes the form
\begin{equation}
\Pi_t(X_\alpha) \simeq e^{-t \phi(x_\alpha)},
\end{equation}
where $x_\alpha=X_\alpha/t$ is the time-intensive photo current per unit time (activity) and $\phi(x_\alpha)$ is the so-called rate function. At very long times, $\Pi_t (X_\alpha)$ converges to a delta-distribution peaked at the mean photo current per unit time $\left<x_\alpha\right>$. For example, this can be easily seen for a Gaussian state, i.e. for a photo current performing a random walk with linear drift. This yields the rate function $\phi(x_\alpha)=(x_\alpha-\left<x_\alpha\right>)^2/2 \Delta x_\alpha^2$, where $\Delta x_\alpha^2$ is the variance of the photo current per unit time, i.e. $\Pi_t (X_\alpha )\sim \mathrm{e}^{-t (x_\alpha-\left<x_\alpha\right>)^2/2 \Delta x_\alpha^2}$. Quantum features, such as negative Wigner functions, can emerge when the rate function differs from a Gaussian shape. However, these can only be uncovered when the interval of the time integration is not too long, i.e. when $\Pi_t (X_\alpha)$ has not yet converged to a delta-distribution \cite{strandberg2019}.

The SCG and the rate functions are related via the Legendre transform:
\begin{equation}
\label{margin}
\phi(x_\alpha) =-\min_s \left[\theta_{X_{\alpha}}(s)+x_{\alpha} s\right].
\end{equation}
The determination of the whole set of distributions \eqref{margin} allows to reconstruct the Wigner distribution of the $X$- and $P$-quadrature activities, $W(x,p)$, using inverting the Radon transform \cite{risken1989}
\begin{eqnarray*}
\Pi_{t=1} (X_\alpha )&\sim& \mathrm{e}^{-\phi(x_\alpha) }\\
&=&\int_\mathbb{R} W\left(x \cos{\alpha}\!-\!p\sin{\alpha},x \sin{\alpha}\!+\! p\cos{\alpha}\right)\mathrm{d}p.
\end{eqnarray*}
In practice, not all $\phi(x_\alpha)$ are measured, but only a discrete set of them. Hence, the WD can be estimated numerically using tomographic imaging software.


\end{document}